# Shape-Margin Knowledge Augmented Network for Thyroid Nodule Segmentation and Diagnosis

Weihua Liu, Chaochao Lin et al.

Beijing Institute of Technology & AthenaEyesCO.,LTD.

**Abstract:** Thyroid nodule segmentation is a crucial step in the diagnostic procedure of physicians and computer-aided diagnosis systems. Mostly, current studies treat segmentation and diagnosis as independent tasks without considering the correlation between these tasks. The sequence steps of these independent tasks in computer-aided diagnosis systems may lead to the accumulation of errors. Therefore, it is worth combining them as a whole through exploring the relationship between thyroid nodule segmentation and diagnosis. According to the thyroid imaging reporting and data system (TI-RADS), the assessment of shape and margin characteristics is the prerequisite for the discrimination of benign and malignant thyroid nodules. These characteristics can be observed in the thyroid nodule segmentation masks. Inspired by the diagnostic procedure of TI-RADS, this paper proposes a shape-margin knowledge augmented network (SkaNet) for simultaneously thyroid nodule segmentation and diagnosis. Due to the similarity in visual features between segmentation and diagnosis, SkaNet shares visual features in the feature extraction stage and then utilizes a dual-branch architecture to perform thyroid nodule segmentation and diagnosis tasks simultaneously. To enhance effective discriminative features, an exponential mixture module is devised, which incorporates convolutional feature maps and self-attention maps by exponential weighting. Then, SkaNet is jointly optimized by a knowledge augmented multi-task loss function with a constraint penalty term. It embeds shape and margin characteristics through numerical computation and models the relationship between the thyroid nodule diagnosis results and segmentation masks. We evaluate the proposed approach on a public thyroid ultrasound dataset (DDTI) and a locally collected thyroid ultrasound dataset. The experimental results reveal the value of our contributions and demonstrate that our approach can yield significant improvements compared with state-of-the-art counterparts.

## 1. Introduction

Thyroid cancer is one of the most common endocrine cancers [34]. Despite a relatively low mortality rate of thyroid cancer, it still threatens the life quality of patients and causes a psychological shadow [1]. Fortunately, early diagnosis and treatment of thyroid nodules can greatly alleviate the deterioration of these diseases and prevent the occurrence of these situations. For early diagnosis of thyroid nodules, ultrasound imaging has become one of the preferred methods due to its noninvasive and safe characteristics [2]. Traditionally, the diagnostic procedure of thyroid nodules is implemented by physicians with their rich clinical experience. With the development of deep learning in natural image analysis, computer-aided diagnosis techniques based on deep neural networks have made significant strides in medical image analysis. For instance, convolutional neural networks (CNN) [5,6] and vision Transformers [7,8] have been widely applied to diagnose thyroid nodules from ultrasound images in recent works. However, accurate diagnosis of thyroid nodules in ultrasound images remains a challenge due to the domain differences between medical images and natural images. In particular, existing computer-aided diagnosis methods are considered unreliable since they lack consideration for specific medical knowledge [3]. Therefore, modeling medical prior knowledge is essential for improving the reliability and interpretability of computer-aided diagnosis systems.

Inspired by the diagnostic procedure of thyroid imaging reporting and data system (TI-RADS) [29], we note the relationship between thyroid nodule segmentation and diagnosis tasks. Generally, a computer-aided diagnosis pipeline mainly consists of detection, segmentation, and diagnosis in sequence. However, the pipeline treats segmentation and diagnosis as independent steps without considering their intrinsic connection. The multiple steps easily lead to the accumulation of errors. In other words, incorrect thyroid nodule segmentation masks may lead to misclassification of the thyroid nodules. Therefore, we integrate thyroid nodule segmentation and diagnosis as a whole, instead of two separate steps. To achieve accurate diagnosis results of thyroid nodules, the segmentation masks of the lesion areas are essential as it provides valuable information. Specifically, the shape and margin characteristics of the predicted segmentation mask are prerequisites for the discrimination of benign and malignant thyroid nodules, as shown in Fig. \ref{fig1}. The thyroid nodule with a well-defined margin and a wider-than-tall shape is likely to be benign, while conversely, it may be malignant. This medical prior knowledge is specified in TI-RADS. The diagnostic procedure of TI-RADS indicates modeling the relationship between the segmentation masks and the diagnosis results is valuable. To our knowledge, there is no previous work that incorporates the medical prior knowledge of shape and margin into the integration of thyroid nodule segmentation and diagnosis.

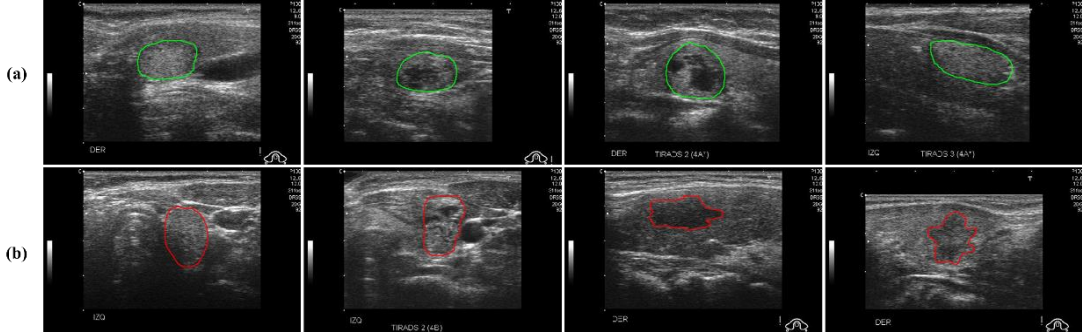

**Figure 1:** The illustration of medical prior knowledge for the discrimination of benign and malignant thyroid nodules. (a) Benign nodules and their segmentation margin (in green). (b) Malignant nodules with their segmentation margin (in red). Note that benign nodules typically exhibit well-defined margins and wider-than-tall shapes, as specified in the TI-RADS.

In this paper, we propose SkaNet, a shape-margin knowledge augmented network for thyroid nodule segmentation and diagnosis. The main contributions of our approach are as follows:

(1) SkaNet is an integrated multi-task architecture. Due to the similarity in visual features between thyroid nodule segmentation and diagnosis, SkaNet shares visual features in the feature extraction stage and utilizes dual branches to perform these tasks simultaneously, which improves diagnosis accuracy and reliability.

(2) In the shared feature extraction stage, SkaNet combines convolutional feature maps and self-attention maps through exponential weighting in a designed exponential mixture module. This process effectively enhances the description of the lesion area and discriminative features.

(3) A knowledge augmented multi-task loss function for joint optimization is defined with a constraint penalty term. The shape-margin knowledge, which is related to the discrimination of benign and malignant thyroid nodules, is embedded in the penalty term through the calculation of the shape and margin characteristics of the segmentation masks.

The remainder of this paper is organized as follows: Section 2 describes related works. Section 3 presents the proposed approach, including our network architecture and the knowledge augmented multi-task learning objective. In Section 4, we discuss the implementation details and present the experimental results. Finally, we conclude in Section 5.

## 2. Related Work

In previous works, thyroid nodule segmentation and diagnosis in ultrasound images were considered independent problems. In this section, we review the methods for thyroid nodule segmentation and diagnosis. Visual attention schema and multi-task learning methods used in the field of thyroid nodule image analysis are also introduced briefly.

### 2.1 Thyroid Nodule Segmentation

Previous thyroid nodule segmentation methods can be divided into two categories: traditional methods and deep learning-based methods. Traditional thyroid nodule segmentation methods include shape-based methods [4] and region-based methods [5].

Different from traditional methods, deep learning-based methods can automatically learn useful features from labeled images to classify the pixels or image blocks. Zhou et al. [8] applied a U-shaped fully convolutional neural network (UNet) [6] and proposed an interactive segmentation method utilizing manual annotation for thyroid nodule segmentation. Wang et al. [9] developed an attention-based semi-supervised neural network with weakly annotated classification data and a small amount of fully annotated segmentation data. Yang et al. [35] utilized two reciprocal networks and the pyramid attention mechanism to segment malignant thyroid nodules. Li et al. [36] developed a Transformer-based feature fusion network, which integrates multi-scale convolutional features and Transformer for better capturing long-range dependencies via the channel and spatial attention mechanism. Gong et al. [31] proposed feature enhancement network for thyroid nodule segmentation, which simultaneously learns nodule size and position.

### 2.2 Thyroid Nodule Diagnosis

Previous thyroid nodule diagnosis methods are based on traditional machine learning techniques and hand-crafted image features. Ozyilmaz et al. [10] proposed a thyroid diagnosis by a multilayer perceptron. Chen et al. [11] constructed a three-stage expert system for diagnosing thyroid disease using a particle swarm optimization SVM method.

Recently, deep learning methods have shown promising applications in thyroid nodule diagnosis. Ma et al. [13] proposed a hybrid method for thyroid nodule diagnosis, which fused two pre-trained convolutional neural networks with different convolutional layers and fully-connected layers. Wang et al. [12] improved the standardized fast region-based convolutional network method for papillary thyroid cancer recognition by connecting shared convolutional layers. Nguyen et al. [27] introduced the analysis of the spatial domain and frequency domain on the basis of CNN to extract more discriminative features for the diagnosis of thyroid nodules. Srivastava et al. [32] established a grid search optimization-based CNN for thyroid nodule identification and classification.

**2.3 Visual Attention Mechanism in Thyroid Nodule Image Analysis**

Visual attention mechanisms have been popular in recent deep learning-based visual analysis. They have been successfully applied in thyroid nodule image analysis. Wang et al. [15] proposed two kinds of attention modules to improve network performance. These modules suppressed or activated the feature channels and image regions respectively through the trainable feedforward structure of bottom-up and top-down. Li et al. [16] introduced a self-attention scheme to extract the features of the dynamically-selected nodule regions ranging from local to global. Wan [14] et al. proposed a novel hierarchical temporal attention network, which leveraged an attention mechanism to embed global enhancement dynamics into each identified salient pattern. Yu et al. [17] proposed an edge-based attention mechanism to strengthen the nodule edge segmentation effect.

**2.4 Multi-task Learning in Thyroid Nodule Image Analysis**

Multi-task learning improves the accuracy of each task by utilizing the correlation between tasks. Recently, it has attracted attention for thyroid nodule image analysis. Song et al. [18] developed a multi-task cascade convolution neural network framework to exploit the context information of thyroid nodules, which integrated thyroid nodule detection and recognition tasks. Zhao et al. [20] proposed a cascade and fusion method of multi-task convolutional neural networks, which was efficient in diagnosing thyroid diseases by segmenting the thyroid region of interest. Yang et al. [19] presented a multi-task cascade deep learning model, which integrated the various domain knowledge in generating higher quality images through quantifying the features of thyroid nodules and used generated ultrasound images to train a support vector machine for thyroid nodule classification. Gong et al. [21] proposed a multi-task learning framework to learn the segmentation of thyroid regions and nodules simultaneously. Deng et al. [22] proposed a multi-task network based on the TI-RADS criteria to associate the outputs of models with the basic descriptors used by clinicians for image interpretation and diagnosis.

**3. The Proposed Approach**

As illustrated in Fig. \ref{fig2}, the proposed SkaNet is composed of two stages for feature extraction and multi-task prediction, respectively. Due to the similarity in visual features between segmentation and diagnosis, SkaNet shares visual features in the feature extraction stage. A CNN sub-network and a Transformer sub-network are combined to obtain shared features. The CNN sub-network extracts high-level convolutional feature maps from the input ultrasound images. It is implemented by a residual U-Net encoder [26], which has shown excellent performance in medical image segmentation tasks. The Transformer sub-network utilizes standard vision Transformer encoders [7] to generate self-attention maps. To capture global and local information, the self-attention maps from Transformer are incorporated with the convolutional feature maps by an exponential mixture module. Specifically, the self-attention maps and convolutional feature maps are mixed by the exponential element-wise dot product. Since the weighted self-attention maps highlight the discriminative features, our network can selectively focus on the informative regions. Then, the mixed feature maps are subsequently fed into the dual branches for multi-task prediction. The segmentation branch takes the mixed feature maps as input and outputs a binary segmentation mask that outlines the thyroid nodule region. The diagnosis branch takes the same input as the segmentation branch and predicts the likelihood of malignancy. Accordingly, our network has two outputs: the thyroid nodule segmentation mask and the diagnosis result. The shape and margin characteristics of thyroid nodules are assessed by the shape and margin assessment module. Furthermore, a constraint penalty term is defined based on the medical prior knowledge of the correlation between these characteristics and diagnosis results. The constraint penalty term is combined with the muti-task loss function for knowledge augmented joint optimization of SkaNet.

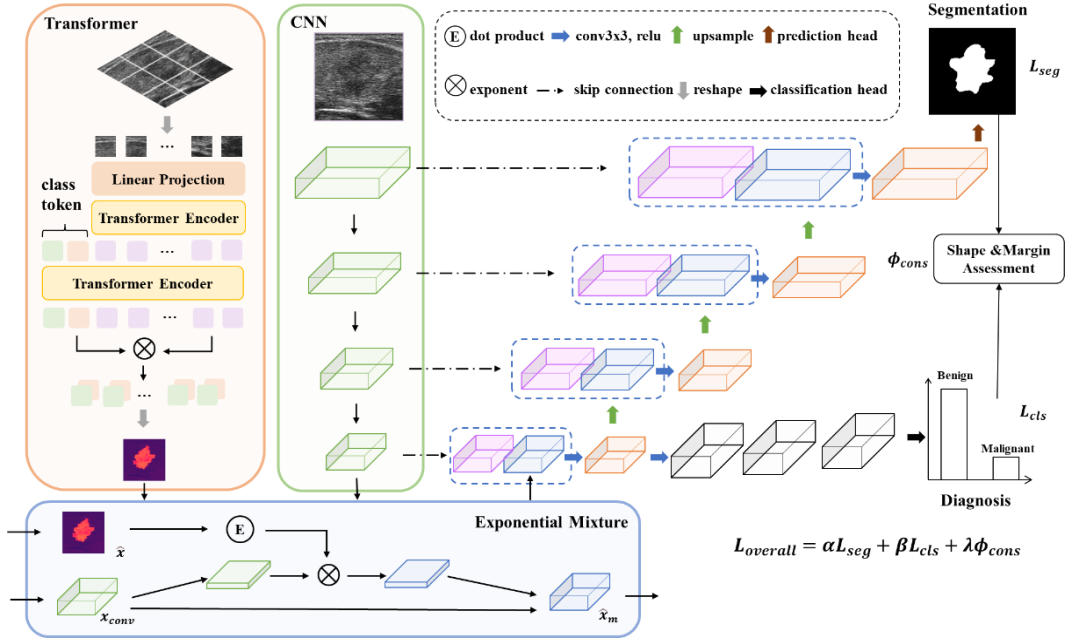

**Figure 2:** The convolutional feature maps and self-attention maps are fused in the exponential mixture module. The mixed feature maps then flow into dual branches to perform thyroid nodule segmentation and diagnosis tasks simultaneously. There are two outputs of SkaNet: the thyroid nodule segmentation mask and the diagnosis result. SkaNet is jointly optimized by a knowledge augmented muti-task loss function with a constraint penalty term $\phi_{cons}$, which is calculated based on the characteristics of the shape and margin assessment module.

### 3.1 Exponential Mixture in Extracted Features

The self-attention mechanism is an efficient method to strengthen the feature extraction of networks. In order to highlight the valuable features for segmentation and diagnosis, our proposed architecture incorporates a self-attention mechanism in the shared feature extraction stage by the exponential mixture of convolutional feature maps and self-attention maps.

Specifically, the input ultrasound images $I \in \mathbb{R}^{H \times W \times 3}$ are partitioned into $N = \frac{H}{P} \times \frac{W}{P}$ patches in the Transformer sub-network, where $P$ is the resolution of each image patch. After the linear projection with output dimension $D_0$ and the combination with a learnable positional embedding, it produces a sequence of patch embeddings $x = \{x_1, x_2, \ldots, x_N\} \in \mathbb{R}^{N \times D_0}$. A Transformer encoder is then utilized to encode patch embeddings and the result is $z = \{z_1, z_2, \ldots, z_N\} \in \mathbb{R}^{N \times D_0}$. Through self-attention computation, the first transformer encoder models the correlation of each ultrasound image patch. To utilize the class prior, two additional tokens $c = \{c_0, c_1\} \in \mathbb{R}^{2 \times D_0}$ are introduced, which represent the class of background and nodule, respectively. These tokens are combined with the encoded image patch embeddings $z$ and fed into the second transformer encoder. The output embeddings are $[c', z'] = [\{c'_0, c'_1\}, \{z'_1, z'_2, \ldots, z'_N\}]$. This process is similar to the vision transformer and is used to encode the correlation between class embeddings and patch embeddings. Finally, the self-attention maps of thyroid nodules $\hat{x}$ are computed as:

$$\hat{x} = \text{Reshape}(z' c'^T_1)$$

where Reshape($\cdot$) represents an operation that reshapes $z' c'^T_1 = \{z'_1 c'^T_1, z'_2 c'^T_1, \ldots, z'_N c'^T_1\} \in \mathbb{R}^{N \times 1}$ into a 2D self-attention map $s \in \mathbb{R}^{\frac{H}{P} \times \frac{W}{P} \times 1}$ and bilinearly upsamples $s$ to $\hat{x} \in \mathbb{R}^{H' \times W' \times 1}$. Note that $H' \times W'$ is the size same as the convolution feature maps from the CNN sub-network.

The obtained self-attention maps are utilized to highlight the thyroid nodules. As described in [28], Transformer has a defect in channel modeling. In order to recalibrate channel-wise feature responses adaptively, the exponential mixture adopts the idea of squeeze and excitation [37] that explicitly models interdependencies between channels. At first, the convolutional feature maps $x_{conv}$ is squeezed into a channel descriptor $x'_{conv}$. It is denoted as:

$$x'_{conv} = \sigma_1(W_1^T x_{conv} + b_1)$$

where $x_{conv} \in \mathbb{R}^{H' \times W' \times C}$ and $x'_{conv} \in \mathbb{R}^{H' \times W' \times 1}$; $W_1 \in \mathbb{R}^{C \times 1}$ is weight of $1 \times 1$ convolution layer; $b_1$ is the corresponding bias; and $C$ refers to input channel sizes of $x_{conv}$. The function $\sigma_1(x) = \max(x, 0)$ corresponds to ReLU activation function.

The self-attention maps of thyroid nodules $\hat{x}$ are then exponentiated and multiplied element-wise with the squeezed convolutional feature maps $x'_{conv}$. The mixing process is defined as

$$\hat{x}' = e^{\hat{x}} x'_{conv}$$

where $\hat{x}' \in \mathbb{R}^{H' \times W' \times 1}$ denotes the squeezed mixed feature maps. With the aid of the exponential operation, the convolutional feature maps are selectively enhanced by the thyroid nodule attention without discarding other features, due to $e^{\hat{x}} \in (0, +\infty)$.

Then, a dimensionality increasing layer is utilized to fully capture channel-wise dependencies. The final mixed feature maps $\hat{x}_m$ are obtained.

$$\hat{x}_m = \sigma_2(W_2^T \hat{x}' + b_2)$$

where $W_2 \in \mathbb{R}^{1 \times C}$ is weight of $1 \times 1$ convolution layer; $b_2$ is the corresponding bias, $\sigma_2(x) = 1/(1 + \exp(-x))$ is the sigmoid activation function. Through the operation of squeeze and excitation in the exponential mixture, the mixed feature maps $\hat{x}_m \in \mathbb{R}^{H' \times W' \times C}$ aggregates the global information and reduces the information noise. $\hat{x}_m$ is fed into the segmentation and diagnosis branch.

## 3.2 Shape and Margin Assessment

**Shape Assessment.** A taller-than-wide shape is a highly specific indicator of malignancy as specified in TI-RADS [29]. This feature can be determined by the aspect ratio (AR) in the axial plane.

$$\text{AR} = \frac{h}{w}$$

where $h$ and $w$ are the longitudinal diameter and transverse diameter of a thyroid nodule in the scanning, respectively. As shown in Fig. 3, the longitudinal diameter refers to the maximum anteroposterior diameter of the thyroid nodule, which is perpendicular to the skin. The transverse diameter refers to the maximum transverse diameter of the thyroid nodule on the horizontal scanning or the axial diameter on the longitudinal scanning, which is parallel to the skin. Intuitively, $h$ and $w$ are the height and width of a thyroid nodule in the ultrasound image.

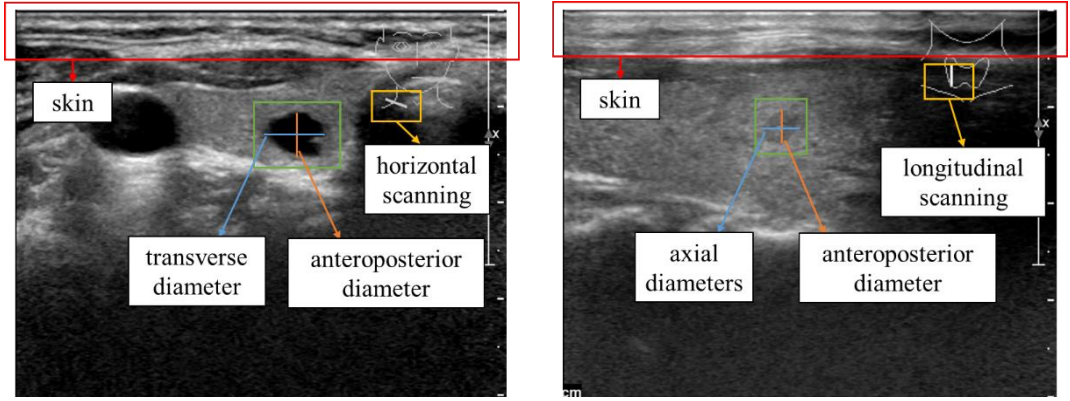

**Figure 3:** Illustration of the computation of the aspect ratio (AR). (a) For the taller-than-wide shape, the longitudinal diameter is greater than the transverse diameter, resulting in an AR value greater than 1; (b) For the wider-than-tall shape, the longitudinal diameter is less than the transverse diameter, leading to an AR value less than 1.

**Margin assessment.** A lobulated or irregular margin, such as a spiculated or jagged edge, is another reliable sign of malignancy as specified in TI-RADS. We aim to incorporate this important medical prior knowledge into our network to augment the reliability and interpretability of the computer-aid diagnosis systems. However, assessing the irregularity of thyroid nodules is usually subjective by physicians and difficult to quantify. Inspired by the Boyce-Clark shape index (BCSI) [29], a commonly used margin measure in cartographic analysis, we propose a novel method for measuring the irregularity of thyroid nodule margin. The main idea of BCSI is to evaluate the margin according to the distance difference between $n$ lines radiating from the center of the graph to the boundary. The BCSI is defined as:

$$\text{BCSI} = \sum_i |\frac{r_i}{\sum_j r_j} - \frac{1}{n}|$$

where $n$ is the number of lines radiating from the center and $r_i$ is the distance from $i$-th center radial to contour margin.

Based on the core idea of the BCSI, we further consider the margin characteristic of thyroid nodules. Obviously, regular thyroid nodules are not necessarily perfectly round, but they will not invade surrounding tissues. In other words, irregular thyroid nodule margins have depressions due to inconsistent invasion rates of cancer cells and normal tissue cells. Therefore, the irregularity (IR) of the margin is calculated based on the convex hull of computational geometry [25] instead of a perfect round. Similar to BCSI, we calculate the distance difference between $n$ lines radiating from the center of the thyroid nodule to the contour margin and its convex hull margin. The IR of a thyroid nodule is denoted as follows:

$$\text{IR}_n = \text{Tanh} \sum_i \frac{R_i - r_i}{R_i}$$

where $n$ is the number of lines radiating from the center, $r_i$ is the distance from the $i$-th center radial to the contour margin, and $R_i$ is the distance from the $i$-th center radial to the convex hull margin, as shown in Fig. 4. The Tanh function is utilized to map values between 0 and 1. Note that the degree of irregularity is bound between 0 and 1, and tend to approach 1 when thyroid nodules are irregular.

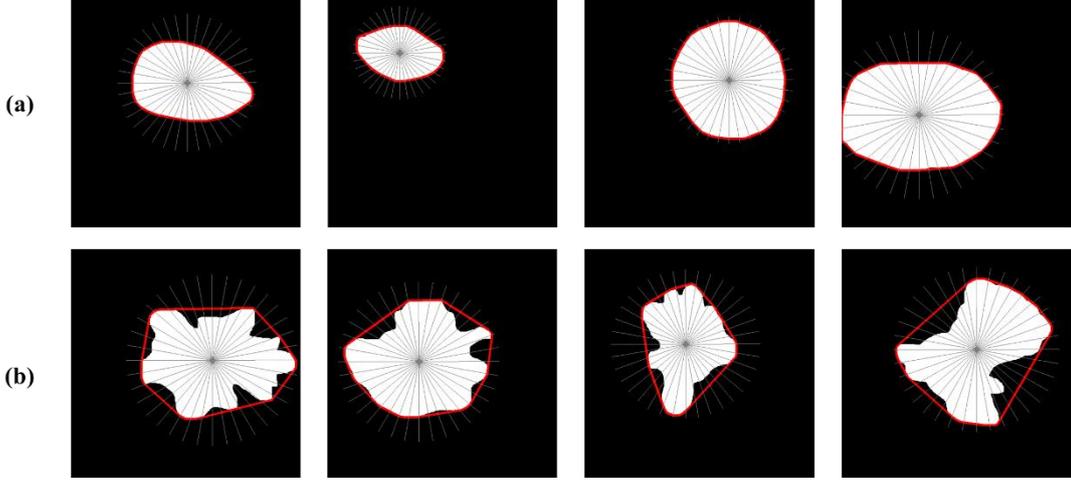

**Fig. 4:** Illustration of the computation of irregularity measure (IR). The convex hulls, center radials and segmentation masks are shown in red, gray, and white, respectively: (a) For regular margin, the distances from the center radials to the contour margin and to the convex hull margin are similar, resulting in an IR value close to 0; (b) For irregular margin, there are significant differences in the these distances, leading to an IR value greater than 0.

**3.3 Knowledge Augmented Multi-task Learning**

The shape and margin assessment imposes quantitatively evaluation of the aspect ratio and irregularity based on the thyroid nodule segmentation masks from the segmentation branch. According to the prior knowledge of the correlation between these characteristics and the diagnosis results, a constraint penalty term $\phi_{cons}$ is designed in multi-task learning. Intuitively, the visual representation of the lesion area should be vaguely consistent with the thyroid nodule properties indicated by the discriminative results from the classification branch. If a thyroid nodule is irregular or in a taller-than-wide shape, it has a higher probability to be malignant.

Specifically, when a thyroid nodule is malignant (positive), there is a high probability that its aspect ratio value is greater than 1. An aspect ratio less than 1 may indicate incorrect predictions. Conversely, for benign (negative) nodules, the opposite is true. $P_{AR(M_i)}$ and $N_{AR(M_i)}$ are defined to model these errors for benign or malignant

thyroid nodules, respectively. They are formulated as:

$$P_{AR(M_i)} = \begin{cases} 1 - AR(M_i), & AR(M_i) < 1 \\ 0, & AR(M_i) \geq 1 \end{cases}, N_{AR(M_i)} = \begin{cases} AR(M_i) - 1, & AR(M_i) \geq 1 \\ 0, & AR(M_i) < 1 \end{cases}$$

where $M_i$ is the $i$-th segmentation mask.

On the other hand, when a thyroid nodule is malignant (positive), there is a high probability that its irregularity value is close to 1. An irregularity value nearly equal to 0 may mean that predictions are incorrect. Therefore, we combine AR and IR with the malignant probability $p$ to obtain the constraint penalty term as follows:

$$\phi_{cons} = \frac{1}{N} \sum_i [p_i(P_{AR(M_i)} + 1 - IR_n(M_i)) + (1 - p_i)(N_{AR(M_i)} + IR_n(M_i))]$$

where $N$ is the total number of training samples in a batch, $n$ is the number of lines radiating from the center in the IR calculation, and $p_i$ is the predicted malignant probability of the $i$-th sample. This constraint penalty term augments the shape-margin knowledge for learning the relationship between thyroid nodule segmentation and diagnosis.

Combining with the above constraint term, the overall multi-task loss function is denoted by the weighted-sum form as follows to jointly optimize the network.

$$L_{overall} = \alpha L_{seg} + \beta L_{cls} + \lambda \phi_{cons}$$

where $\alpha, \beta$ and $\lambda$ are the weights for segmentation loss $L_{seg}$, classification loss $L_{cls}$ and constraint penalty term $\phi_{cons}$, respectively. This knowledge augmented multi-task learning improves the reliability of the diagnosis results.

For the diagnosis branch, the classification loss $L_{cls}$ is a standard cross-entropy loss, which is defined as follows:

$$L_{cls} = \frac{1}{N} \sum_i -[y_i \cdot \log(p_i) + (1 - y_i) \cdot \log(1 - p_i)]$$

where $N$ is the total number of training samples in a batch, $y_i$ is the ground truth of the $i$-th sample, and $p_i$ is the predicted probability of the $i$-th sample.

The segmentation loss $L_{seg}$ for segmentation branch is a combination of dice loss and cross-entropy loss, which is formulated as follows:

$$L_{seg} = \frac{1}{2}(1 - \frac{2 \sum_i \sum_c g_i^c s_i^c}{\sum_i \sum_c g_i^{c2} + \sum_i \sum_c s_i^{c2}} + \frac{1}{N} \sum_i \sum_c g_i^c \log s_i^c)$$

where $N$ is the total number of training samples in a batch, $g_i^c$ is the ground truth of class $c$ in the $i$-th sample and $s_i^c$ is the predicted probability of class $c$ in $i$-th sample.

To obtain satisfactory nodule detection results, we should balance these three objectives. In the beginning stage of learning, we should focus more on the thyroid nodule segmentation task because it is a prerequisite step. When thyroid nodule segmentation goes well, we should focus on thyroid nodule diagnosis and knowledge augment. Therefore, we start with $\alpha$ larger than $\beta$ and $\lambda$. Then, when $L_{seg}$ oscillates such that the magnitude of the changes of $L_{seg}$ in sufficient iterations is very small, we reduce $\alpha$ to a value smaller than $\beta$ and $\lambda$. Furthermore, considering that the constraint penalty term is an auxiliary knowledge augment, $\beta$ should be larger than $\lambda$. We provide an example of the parameter settings for balanced coefficients in the experiments.

## 4 Experiments

### 4.1 Experimental Setting

The Digital Database of Thyroid Ultrasound Images (DDTI) [23], which is an open access database, is adopted to evaluate our method. It consists of 347 sets of thyroid ultrasound images from 299 patients with complete annotations and diagnostic descriptions of suspicious thyroid lesions by expert radiologists in XML files. The other dataset is a local dataset, consisting of 17 patients and 458 images to further verify the effectiveness of the main contribution of SkaNet. The evaluation on these two datasets is conducted independently. For each dataset, five-fold cross-validation experiments are conducted. In each round of experiments, we take eighty percent of the ultrasound images as the training set and the remaining images as the test set. The reported results in the following are the means across the five folds.

In our experiments, we use Accuracy (ACC), Specificity (SPEC), Sensitivity (SENS) and F1-score (F1) to evaluate the thyroid nodule diagnosis performance. They are widely used for evaluating the diagnosis. The formulas for ACC, SPEC, SENS and F1 are as follows:

$$ACC = \frac{TP + TN}{TP + FP + TN + FN}$$
$$SPEC = \frac{TN}{TN + FP}$$
$$SENS = \frac{TP}{TP + FN}$$
$$F1 = \frac{2 \times precision \times recall}{precision + recall}$$

where TP, FP, TN and FN are donated as the number of instances correctly diagnosed as malignant, incorrectly diagnosed as malignant, correctly diagnosed as benign and incorrectly diagnosed as benign, respectively. To evaluate the thyroid nodule segmentation performance, we use Intersection-over-Union (IoU) and Dice Score (DSC), as they are popular image segmentation quality criterions:

$$IoU = \frac{|S \cap T|}{|S \cup T|}$$
$$Dice = \frac{2|S \cap T|}{|S| + |T|}$$

where $S$ and $T$ are the predicted and ground-truth thyroid nodule segmentation masks, respectively. The average IoU and DSC for all the ultrasound images during the testing phase are used as our measure of thyroid nodule segmentation quality.

**4.2 Implementation Details**

We resize images and annotations to 512x512 pixels. The scaling ratios of height and width are recorded for each ultrasound image. These scaling ratios are used to compute the original AR and IR from the segmentation mask. Downsampling and upsampling are performed via strided and transposed convolution operations, respectively. We always perform down/up sampling by a factor of 2. The encoder of the CNN sub-network includes four residual convolution blocks. Thus, the size of the feature map from the feature extraction stage is set to $H/16 \times W/16$, where $H$ and $W$ are the width and height of the input image, respectively. The parameters of Transformer sub-network are set to a patch size of 16x16, 3 layers, a 64-dimensional embedding and 2 heads. The decoder of the segmentation branch includes four deconvolution blocks for upsampling to the resolution of the input image.

The parameters in our approach are set as follows and according to experience. For the balanced coefficients in knowledge augmented multi-task learning, we set ($\alpha =$

1.0, $\beta = 0.2$, $\lambda = 0.1$) and change the to ($\alpha = 0.01$, $\beta = 2.0$, $\lambda = 1.0$) when $L_{seg}$ oscillates. If the average difference between each pair of losses in two adjacent epochs is below 0.02, we determine that $L_{seg}$ is in the oscillation state.

Adam optimizer is applied. Our network is trained for 400 epochs with a weight decay of $10^{-4}$, a momentum of 0.9 and a batch size of 4. The learning rate is initialized at $10^{-4}$ and then divided by 10 after every 100 epochs.

### 4.3 Experiments Results

### 4.3.1 Effectiveness of Our Approach

Our approach achieves a Dice of 0.8498 for thyroid nodule segmentation and an accuracy of 98.06% for thyroid nodule diagnosis in the DDTI dataset, while a Dice of 0.8601 and an accuracy of 94.61% in the local dataset. Examples of thyroid nodule segmentation and diagnosis results are shown in Fig. 5. This figure demonstrates that our approach achieves good thyroid nodule diagnosis and segmentation results. These two subtasks are improved with each other. As shown in Fig. \ref{fig5}, our method performs accurate discrimination of the benign and malignant thyroid nodules based on the assistance of a segmentation mask. Interestingly, it can be observed from the predicted thyroid nodule areas of some samples that calcification lesions (strong echoic spots) are distinguished with a lower value, although this information is not annotated in the segmentation ground-truth. The echogenic foci (including macrocalcifications, peripheral calcifications, and punctate echogenic foci) are the other basis for the diagnosis of benign and malignant thyroid nodules, as specified in TI-RADS [29]. This phenomenon indicates that due to the correlation between segmentation and diagnosis tasks, the SkaNet also learns knowledge about different echo intensities in the thyroid nodule area and expressed it in the predicted mask. Furthermore, the nodule areas and their margin have higher responses than normal tissue areas in the class activation maps as shown in Fig. 6. The high response areas represent the attentive content of the diagnosis branch, making the diagnosis results easy to interpret and understand. This finding indicates that not only is thyroid nodule segmentation helpful for thyroid nodule diagnosis but also thyroid nodule diagnosis affects the thyroid nodule segmentation results. We compare the results of our approach with those of other methods, including TRFE+ [31], FCG-Net [38], TFNet [28], ResNet+Inception [27], MHDL [33], and GSO-CNN [32]. Note that the results of these compared methods are reported in their paper. The comparison results are shown in Table \ref{tab1}. The results demonstrate that the proposed approach stably outperforms comparison methods by a wide margin. Our approach leads to better accuracy, specificity, sensitivity, and F1 score. The corresponding increases in the ACC are 6.01%, 5.05%, and 2.76% compared with ResNet+Inception, MHDL, and GSO-CNN, respectively. In terms of thyroid nodule segmentation, our method surpasses TRFE+ and FCG-Net on Dice by 0.0961 and 0.0456, respectively. Even without a special design for segmentation tasks, our method can slightly exceed the best segmentation result by TFNet (a Dice of 0.8460). This performance indicates that knowledge augmented multi-task learning is reliable. By analyzing the errors from our approach, we find that the predicted segmentation mask for false positive nodules usually has an irregular margin while the margin of false negative nodules is relatively regular. Some examples of these cases of errors are shown in Fig. 7. This highlights the fact that shape and margin are only a part of the medical prior knowledge and the diagnosis of these difficult cases requires additional information. A possible way to deal with these difficult cases is by incorporating more medical prior knowledge in the training stage of the network. We will explore this issue in future works.

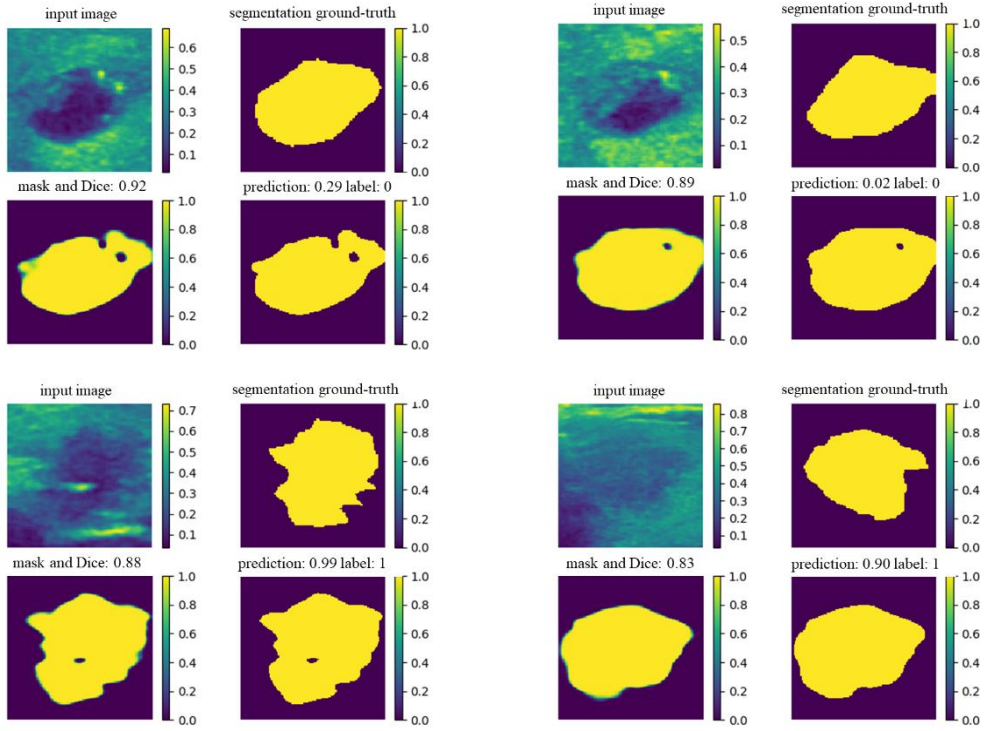

**Figure 5:** Examples of the thyroid nodule segmentation and diagnosis results. There are four groups of results. Two of these groups are classified as benign, and the other two groups are malignant. Each group includes an input image, a segmentation ground-truth, a predicted segmentation mask before argmax operation and a predicted segmentation mask after argmax operation. Their Dice, predicted malignancy probability, and labels are displayed in values (1 indicates a malignant thyroid nodule while 0 indicates a benign thyroid nodule).

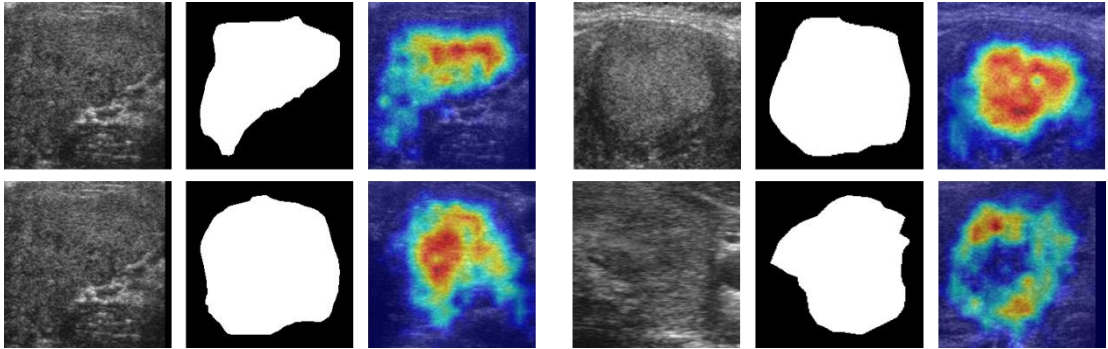

**Figure 6:** Visualization of class activation maps (CAMs) [30] generated by the diagnosis branch of SkaNet.

**Table 1:** The comparison of ACC, SPEC, SENS, F1, IoU and DSC with state-of-the-art methods.

| Method | ACC | SPEC | SENS | F1 | IoU | DSC |
|---|---|---|---|---|---|---|
| TRFE+ [31] | / | / | / | / | 0.6047 | 0.7537 |
| FCG-Net [38] | / | / | / | / | 0.6725 | 0.8042 |
| TFNet [28] | / | / | / | / | 0.7331 | 0.8460 |
| ResNet+Inception [27] | 0.9205 | 0.6569 | 0.9607 | 0.9485 | / | / |
| MHDL [33] | 0.9301 | 0.8679 | 0.9601 | 0.9380 | / | / |
| GSO-CNN [32] | 0.9530 | 0.9487 | 0.9666 | 0.9720 | / | / |
| SkaNet(ours) | **0.9806** | **0.9750** | **0.9826** | **0.9869** | **0.7388** | **0.8498** |

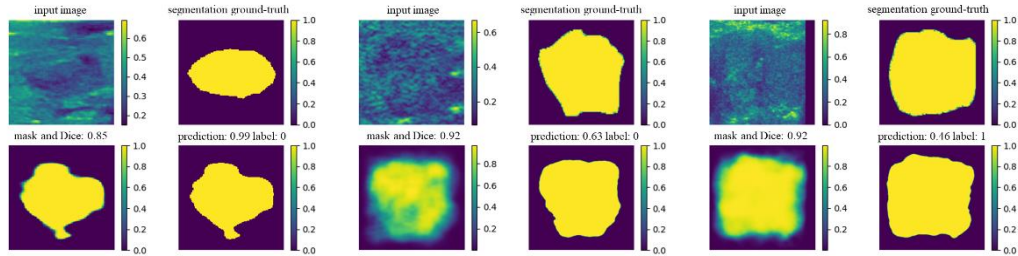

**Figure 7:** Example of failure cases by our approach. The first and second groups are false positive nodules, while the third group is a false negative nodule.

### 4.3.2 Ablation study

Three new ideas in this paper are integrating thyroid nodule segmentation and diagnosis, exponential mixture with self-attention maps, and the knowledge augmented constraint penalty term for multi-task learning. We study the effectiveness of each of the three contributions in this section. Accordingly, we consider three configurations of ablated networks. The three configurations are as follows: without thyroid nodule segmentation (w/o TS), without exponential mixture (w/o EM), and without knowledge augmented constraint penalty term (w/o KA). It should be noted that the configuration w/o TS is optimized only by classification loss. We conduct five-fold cross-validation experiments for each configuration. The corresponding mean performance metrics for each configuration are listed in Table 2.

**Table 2:** Performance results (ACC, SPEC, SENS, F1, IoU, and DSC) from the different configurations of components on DDTI and local dataset.

| Dataset | Configurations | ACC | SPEC | SENS | F1 | IoU | DSC |
|---|---|---|---|---|---|---|---|
| DDTI | w/o TS | 0.9258 | 0.8875 | 0.9391 | 0.9494 | / | / |
| | w/o EM | 0.9581 | 0.9375 | 0.9652 | 0.9716 | 0.6812 | 0.8104 |
| | w/o KA | 0.9451 | 0.9250 | 0.9522 | 0.9626 | 0.7073 | 0.8286 |
| | full SkaNet | **0.9806** | **0.9750** | **0.9826** | **0.9869** | **0.7388** | **0.8498** |
| Local | w/o TS | 0.8876 | 0.8210 | 0.9057 | 0.9269 | / | / |
| | w/o EM | 0.9210 | 0.8737 | 0.9340 | 0.9489 | 0.7236 | 0.8397 |
| | w/o KA | 0.9101 | 0.8632 | 0.9229 | 0.9417 | 0.7338 | 0.8465 |
| | full SkaNet | **0.9461** | **0.9052** | **0.9571** | **0.9653** | **0.7545** | **0.8601** |

According to the results shown in Table 2, we can conclude that:

(1) Thyroid nodule segmentation is important for the success of diagnosis. Compared with the configuration without thyroid nodule segmentation, the embedding of the thyroid nodule segmentation module contributes to obvious improvements in diagnosis accuracy both on DDTI and the local dataset. This phenomenon further demonstrates the feasibility of combining the thyroid nodule segmentation and diagnosis tasks.

(2) The proposed exponential mixture with self-attention maps is effective.

Benefiting from the exponential mixture, the self-attention maps are used in a soft attention manner instead of a hard classification manner, which enables better extraction of global contexts. The corresponding diagnosis accuracy rates of increase are 2.25% and 2.51% on DDTI and the local dataset, respectively.

(3) The constraint penalty term for knowledge augmented learning further increases the Dice of thyroid nodule segmentation and accuracy of diagnosis, from 94.51%/0.8286 to 98.06%/0.8498 and 91.01%/0.8465 to 94.61%/0.8601 on DDTI and the local dataset, respectively. It illustrates that incorporating shape-margin knowledge into a constraint penalty is effective and beneficial to guide the learning of the relationship between thyroid nodule segmentation and diagnosis.

(4) Compared to ablated models, our full model SkaNet achieves higher accuracy in both thyroid nodule segmentation and diagnosis, which shows that our multi-task architecture and knowledge augmented learning is more suitable for thyroid nodule diagnosis in ultrasound images.

## 5  Conclusion

This study proposes SkaNet, a shape-margin knowledge augmented network for thyroid nodule segmentation and diagnosis. Our approach integrates thyroid nodule segmentation and diagnosis by shared feature extraction and dual branches for these tasks. To better extract features of the lesion area, SkaNet utilizes an exponential mixture module with a weighting operator to fuse convolutional feature maps and self-attention maps. More importantly, considering the relationship between thyroid nodule segmentation and diagnosis, a shape-margin knowledge augmented learning method is devised to optimize SkaNet jointly. Specifically, by assessing the shape and margin characteristics of the thyroid nodule segmentation mask, a constraint penalty term is introduced into the multi-task loss function, which embeds the shape-margin knowledge about the discrimination of benign and malignant thyroid nodules. Experimental results on DDTI demonstrate that our approach outperforms state-of-the-art methods. The ablation study on DDTI and the local dataset further confirms the effectiveness of our main contributions. Despite the success of shape-margin knowledge augmented learning, it still has limitations for difficult samples because shape-margin knowledge is only part of medical diagnostic knowledge. In the future, we plan to extend our approach to incorporate more medical knowledge by numerical calculation and apply knowledge augmented learning to improve the reliability of the computer-aid diagnosis systems for wider tumor diagnosis problems.